\documentclass[11pt]{article}
\usepackage {graphicx}
\usepackage {color}
\begin{document}

\title{Hyperbolic and non-hyperbolic chaos in a pair of coupled \\ alternately excited FitzHugh-Nagumo systems}

\maketitle

\begin{center}
Alexey Yu. Jalnine \footnote{{\it Electronic address}: Jalnine@rambler.ru}

\bigskip

{\it Saratov Branch of Kotel'nikov's Institute of Radio-Engineering and Electronics of RAS, \\ Zelenaya 38, Saratov, 410019, Russia}

\end{center}

\begin{abstract}
We investigate a possibility of realization of structurally stable chaotic dynamics in neural systems. The considered model of interacting neurons consists of a pair of coupled FitzHugh-Nagumo systems, with the parameters being periodically modulated in antiphase, so that the neurons undergo alternate excitation with a successive transmission of the phase of oscillations from one neuron to another. It is shown that 4D map arising in a stroboscopic Poincar\'e section of the model flow system possesses a hyperbolic strange attractor of the Smale-Williams type. The dynamical regime observed in the system represents a sequence of amplitude bursts, in which the phase dynamics of oscillatory spikes is described by chaotic mapping of Bernoulli type. The results are confirmed by numerical calculation of Lyapunov exponents and their parameter dependencies, as well as by direct computation of the distributions of angles between stable and unstable tangent subspaces of chaotic trajectories.

\vspace{3mm}


\end{abstract}

\section{Introduction}
\label{sec:Int}

In recent decades, sufficient progress has been achieved in investigations of biophysical and neurophysiological systems via methods of physics and nonlinear dynamics \cite{Wil,Izh,Hil,CFL,VTA}. A close attention was paid both to immediate investigation of natural neural systems using methods of analysis of observed time series \cite{CFL,VTA,MAH,AM,MAA,ANM,HJA,GVC}, and to analysis of the models of neurophysiologic systems described by differential equations founded on specific character of physicochemical processes and neural activities (Hodgkin-Huxley \cite{HH}, FitzHugh-Nagumo \cite{FHN} and Hindmarsh-Rose \cite{HmR} models). Basing on these studies, it is known that both individual neurons and their coupled ensembles can demonstrate complex and chaotic dynamics, bifurcations, a variety of synchronization phenomena and transitions order-to-chaos \cite{CFL,VTA,MAH,AM,MAA,ANM,HJA,AMI,FNP,PVV,SCC,RA}. In particular, immediate studies of the chaotic dynamics of real neurons and their models are presented in the works \cite{CFL,VTA,MAH,AM,MAA,HJA,RA}.

The present paper is devoted to the problem of manifestation of {\it robust} (i.e. structurally stable) chaotic dynamics by the systems of neural nature. In itself, the question about existence of robust chaos in real physical systems is a fundamental one and goes back to the moment of the discovery of a {\it hyperbolic strange attractor} \cite{Ott,SPKuzBook}. Such attractors consisting of hyperbolic saddle chaotic trajectories with well-defined stable and unstable directions possess a remarkable property of structural stability: their dynamical and metrical properties do not change under variations of parameters and functions in dynamical equations, at least while these variations are not too large \cite{SPKuzBook,HY}. In particular, there are no bifurcations of the attractors, and the Cantor-like structure of the fractal attracting set persists under variation of parameters, their Lyapunov exponents demonstrate smooth dependence upon parameters or even remain unchanged, etc. This property sets them apart from ``usual'' chaotic systems (as excited and self-oscillating nonlinear systems of R\"{o}ssler, Duffing, Chua etc.) demonstrating rather different behaviour: they  undergo bifurcations under infinitesimally small parameter variations and perturbations of functions in dynamical equations \cite{AAN}. Such objects demonstrating non-smooth dependencies of dynamical and  metrical characteristics upon the driving parameters are called as ``quasiattractors'' \cite{ABS}. Note that the search of a ``true'' hyperbolic chaotic attractor in realistic systems was never ceased. In application to neurophysiology, one should take a note of the work by Belykh and co-authors \cite{BBM}, in which a theoretical possibility of realization of the hyperbolic attractor of Plykin type for Hindmarsh-Rose \cite{HmR} neuron model was argued. However, there were no concrete results with explicit form of equations and parameter values and their numerical analysis presented in this work. The question about existence of hyperbolic chaos in physical systems and their realistic models remained undisclosed until the appearance of the pioneering work of Kuznetsov \cite{SPKuz1} in which a simple and physically realizable example of uniformly hyperbolic chaotic attractor of the Smale-Williams type was presented for the system of two coupled alternately excited van der Pol oscillators, and subsequent works (see \cite{SPKuzBook} and references therein).

In the present paper, a possibility of realization of hyperbolic chaos in neural system is manifested. We consider a pair of coupled FitzHugh-Nagumo systems as a model of two interacting neurons. The control parameters responsible for excitation in each system undergo slow periodic modulation in time. This modulation is shifted for a half period between the neurons, so that on a half-period the first neuron is excited and the second one is below the threshold of generation, while on the another half-period the situation is the opposite. The coupling allows to transmit the phase of excitation from one neuron to another. It is shown that, if the parameter values and the mode of interaction of the neurons are chosen appropriately, the neurons generate periodic sequences of ``bursts'', in which the phase dynamics of the oscillations (``spikes'') is described by a simple chaotic 1D map of ``double'' Bernoulli type ($\varphi' =4\varphi \pmod{2\pi}$). In terms of the Poincar\'e map obtained by stroboscopic section of the flow, the attractor of the whole system is of the same kind as the Smale-Williams solenoid, but embedded in a 4D state space (instead of a 3D space, as in the original Smale-Williams model \cite{SW}).

\section{The dynamics of two coupled FitzHugh-Nagumo neurons}
\label{sec:Dynamics}

The FitzHugh-Nagumo equations can be written as $\dot{x}=cx-x^3/3-y$, $\dot{y}=ax-by+I$, where $x$ and $y$ signify the membrane potential and the slow recovery variable, respectively, $a$, $b$ and $c$ are parameters (constants in the original model), and the parameter $I$ is the external current across the membrane. It is known that the FitzHugh-Nagumo system can be reduced to the van der Pole equation with additional cubic nonlinearity via simple variable and  parameter change. Therefore, we can use the same principles as \cite{SPKuz1} when constructing our model system. Let us consider the pair of coupled FitzHugh-Nagumo systems in the following form:
\begin{equation}
I: \ \left\{
\begin{array}{lll}
\dot{x}=c_1x-x^3/3-y, \\
\dot{y}=a_1x-b_1y+\varepsilon \dot{u}^2,
\end{array}
\right.
\ \ \
II: \ \left\{
\begin{array}{lll}
\dot{u}=c_2u-u^3/3-v, \\
\dot{v}=a_2u-b_2v+\varepsilon \dot{x}^2.
\end{array}
\right.\label{eq:e1}
\end{equation}
The pairs of the variables $(x,y)$ and $(u,v)$ relate to the first and the second subsystems, respectively. The strength of coupling is characterized by $\varepsilon$. The coupling is symmetric and depends upon the squared velocity of the membrane potential change (i.e., upon the membrane current, rather than upon the membrane charge). Let the coefficients $a_{1,2}$ and $b_{1,2}$ (whose ratio is responsible for the Hopf bifurcation in each of the uncoupled subsystems) undergo slow harmonic modulation in time with the frequency $\Omega$:
\begin{equation}
\begin{array}{ll}
a_{1,2}(t)=A_{0} \pm A_{1}\sin{\Omega t},
b_{1,2}(t)=B_{0} \pm B_{1}\sin{\Omega t}.
\end{array}
\label{eq:e2}
\end{equation}
The variation of the pairs $(a_{1},b_{1})$ and $(a_{2},b_{2})$ occurs in a counter-phase, so that the subsystems undergo excitation alternately, the either one on its' own half of the period $T=2\pi/\Omega$.

In order to proceed to the complete 4D map with a hyperbolic attractor of Smale-Williams type, we perform a procedure of stroboscopic section of the flow system~(\ref{eq:e1}) at discrete time moments $t_n=t_0+nT$  ($n$ is integer). Supposing that instantaneous state of the system~(\ref{eq:e1}) is given by a vector ${\bf Z}_{n}=(x(t_n),y(t_n),u(t_n),v(t_n))$, the evolution of ${\bf Z}_{n}$ on one period of parameters modulation is described by a 4D vector function ${\bf F}({\bf Z})$ operating in the space of vectors ${\bf Z}$:
\begin{equation}
{\bf Z}_{n+1}={\bf F}({\bf Z}_{n}).
\label{eq:e3}
\end{equation}
This procedure defines a Poincar\'e map for our system and provides an alternative description of the dynamics in terms of discrete-time states. Technically, the construction of the map~(\ref{eq:e3}) may be performed by numerical solution of the equations~(\ref{eq:e1}) with the conditions~(\ref{eq:e2}).

In order to understand the dynamics of the system~(\ref{eq:e1}) on a qualitative level, let us suppose that on a stage of spikes generation the first neuron has some phase $\varphi$: $x \sim \cos{(\omega t+\varphi)}$, where $\omega$ is a characteristic frequency of spikes. The squared value $\dot{x}^2$ contains the second harmonic: $\sin{(2\omega t+2\varphi)}$, and its phase is $2\varphi$. Following the idea of the work \cite{SPKuz1}, one can expect that, as the first half-period comes to the end, and the second neuron becomes excited, the induced oscillations of the variable $u$ inherit the same phase $2\varphi$: $u \sim \sin{(\omega t+2\varphi)}$. Note that the mechanism of phase transmission has to be non-resonant in this case. On the next half-period the situation is opposite, and the second neuron affects on the first one via quadratic term $\dot{u}^2 \sim \cos{(2\omega t+4\varphi)}$, so that the first neuron accepts the phase $4\varphi$. Thus, the phase of oscillations undergoes doubling twice during one period of the parameters modulation, when it is transmitted from one neuron to another and backwards. Then, the initial phase becomes quadrupled during one period $T$, so that the evolution of the phase taken at discrete time moments $t_n$ is described by the double Bernoulli map
\begin{equation}
\varphi_{n+1}=4\varphi_n \pmod{2\pi}.
\label{eq:e4}
\end{equation}

\begin{figure}[htbp]
\centerline{\includegraphics[width=6.5in]{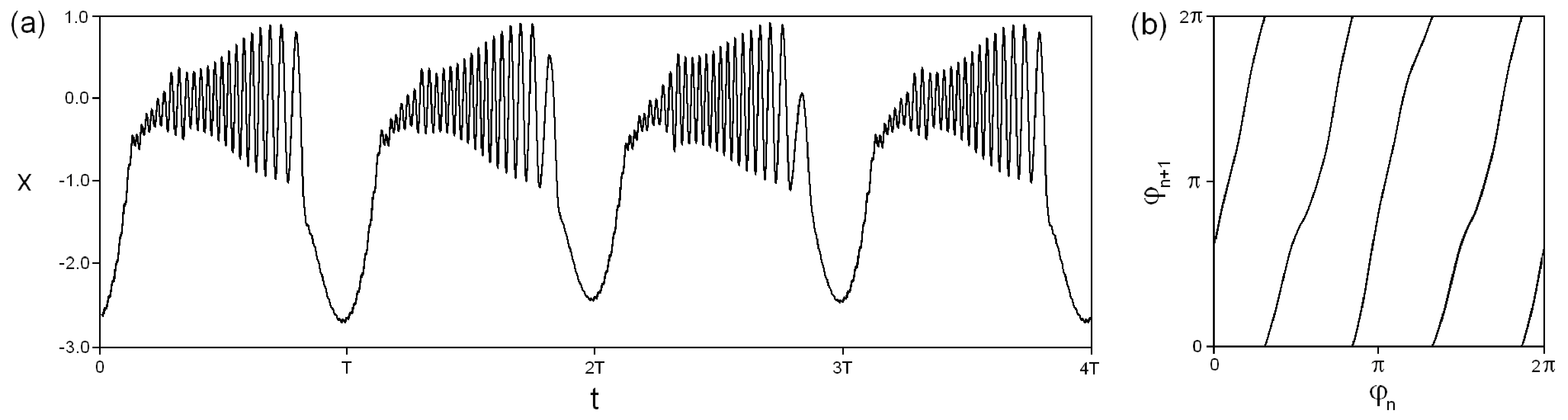}}
\caption{A typical fragment of time realization of chaotic bursts (a) and an empirically obtained map for the phase variable (b) for the system~(\ref{eq:e1}) at the values of parameters given in the text.}
\label{fig:f1}
\end{figure}

Fig.~\ref{fig:f1}(a) shows a fragment of time dependence for the variable $x$ of the system~(\ref{eq:e1}), where the coefficients $a_{1,2}$ and $b_{1,2}$ are modulated in accordance with~(\ref{eq:e2}), at the parameter values  $A_0=1.5$, $A_1=1.7$, $B_0=0.1$, $B_1=0.1$, $c=0.2$, $\Omega=0.05$, $\varepsilon=0.7$. The plot shows a characteristic pattern of neural bursts, i.e. sequences of quick oscillations of the membrane potential variable (spikes), alternating with stages of slow damping/recovery of activity. The bursts amplitude is modulated periodically, due to the supposed periodic modulation of the parameters.
An empirical diagram of the return map for the phase of the first neuron is shown in the Fig.~\ref{fig:f1}(b). The phases are determined at discrete time moments $t_n=t_0+nT$ on the stages of spikes generation via formula $\varphi=\arg{(x-i\dot{x})}$. As expected, the map is close to the relation~(\ref{eq:e4}):
\begin{equation}
\varphi_{n+1}=4\varphi_n +f(\varphi_n) \pmod{2\pi},
\label{eq:e5}
\end{equation}
where $f(\varphi)$ is a $2\pi$-periodic smooth nonlinear function. This map is topologically equivalent to the double iterated chaotic Bernoulli map; it differs slightly because of the present nonlinear part, which arises due to imperfection of the above qualitative arguments.

The complete 4D Poincar\'e map~(\ref{eq:e3}) appearing in the stroboscopic section of the system~(\ref{eq:e1}) must have one expanding Lyapunov direction and three contracting ones. The unstable direction is associated with the phase variable $\varphi$  and is characterized by a positive Lyapunov exponent $\Lambda_1 \approx \ln{4}$. Three other directions are stable and correspond to a 3D stable manifold of the attractor. The respective Lyapunov exponents are negative ($\Lambda_{2,3,4}<0$). Interpreting the effect of the mapping upon a domain in the phase space volume enclosing the attractor, we can image a 4D toroid (which is a direct product of a 3D ball and a 1D circle) and associate one iteration of the map with longitudinal stretch, contraction in the transversal directions, and insertion of the four times folded ``tube'' into the original toroid.

\begin{figure}[htbp]
\centerline{\includegraphics[width=2.5in]{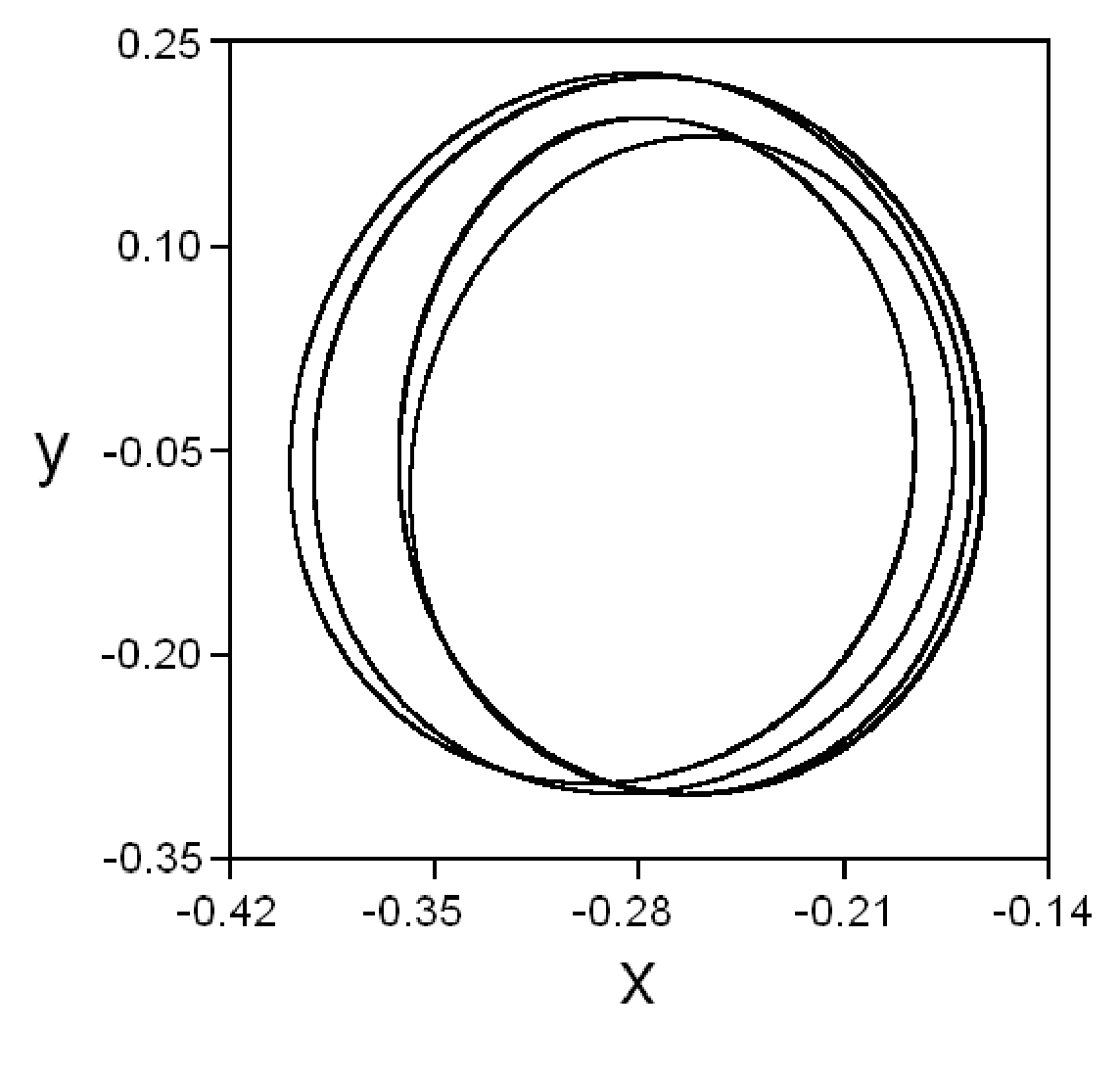}}
\caption{A portrait of the strange attractor of Smale-Williams type in the stroboscopic Poincar\'e section in projection onto the plane $(x,y)$.}
\label{fig:f2}
\end{figure}

Fig.~\ref{fig:f2} shows a projection of stroboscopic section of the attractor of the map on the $(x,y)$ plane (parameter values are taken the same as in the Fig.~\ref{fig:f1}). It looks like the solenoidal Smale-Williams attractor, i.e. represents a ``thread'' with infinite number of turns, and possesses a Cantor-like structure, which can be seen in high resolution. Running ahead, we may say that an estimation of fractal dimension of the attractor via the Kaplan-Yorke \cite{Sch} formula (with above selected parameters of the system) gives $D_{\Lambda} \approx 1.17$.
To characterize the chaotic regime in the system~(\ref{eq:e1}) quantitatively, we calculated numerically the full spectrum of Lyapunov exponents using the Benettin algorithm \cite{BGG}. For that, we linearize equations~(\ref{eq:e1}) and obtaine a system in variations:
\begin{equation}
I: \ \left\{
\begin{array}{lll}
\dot{\tilde{x}}=c_1\tilde{x}-x^2\tilde{x}-\tilde{y}, \\
\dot{\tilde{y}}=a_1\tilde{x}-b_1\tilde{y}+2\varepsilon\dot{u}\dot{\tilde{u}},
\end{array}
\right.
\ \ \
II: \ \left\{
\begin{array}{lll}
\dot{\tilde{u}}=c_2\tilde{u}-u^2\tilde{u}-\tilde{v}, \\
\dot{\tilde{v}}=a_2\tilde{u}-b_2\tilde{v}+2\varepsilon\dot{x}\dot{\tilde{x}}.
\end{array}
\right.
\label{eq:e6}
\end{equation}
It describes evolution in time for a perturbation vector $(\tilde{x}(t), \tilde{y}(t), \tilde{u}(t), \tilde{v}(t))$ at some trajectory of the original system~(\ref{eq:e1}). For calculation of the full spectrum of Lyapunov exponents, the system~(\ref{eq:e1}) was integrated numerically simultaneously with a collection of four linearized systems~(\ref{eq:e6}) with different values of initial perturbations $(\tilde{x_k}(t_0),\tilde{y_k}(t_0),\tilde{u_k}(t_0),\tilde{v_k}(t_0))$, $k=1 \ldots 4$. The integration was executed over time interval $NT$, where $N \cong 10^4$. In the course of integration, at each next moment $t_n=t_0+nT$, $n=1 \ldots N$, the Gram-Schmidt orthogonalization and normalization was performed for vectors $\tilde{{\bf Z}}_{k}(t_n) = (\tilde{x_k}(t_n), \tilde{y_k}(t_n), \tilde{u_k}(t_n), \tilde{v_k}(t_n))$, $k=1 \ldots 4$. At the same time, we calculated the mean rates of growth or decrease of the accumulated sums of logarithms of the norms $S_k=\sum_{n=1}^{N}\ln{|\tilde{{\bf Z}}_k'(t_n)|}$ (where $\tilde{{\bf Z}}_k'$ is a perturbation vector after the orthogonalization, but before the normalization). Then, the Lyapunov exponents of the map~(\ref{eq:e3}) may be estimated as $\Lambda_k=S_k/N$, while the exponents of the original system~(\ref{eq:e1}) are linked to them as $\lambda_k=\Lambda_k/T$. For the attractor presented in the Fig.~\ref{fig:f2}, the numerically obtained Lyapunov exponents are $\Lambda_1 \approx 1.37$, $\Lambda_2 \approx -7.76$, $\Lambda_3 \approx -17.24$, $\Lambda_4 \approx -299.0$. As expected, the first exponent appears to be roughly $\Lambda_1 \approx \ln{4}$, while other ones are strongly negative. This fact completely agrees with our qualitative arguments concerning the expected values of the Lyapunov exponents for the attractor of Smale-Williams type.

\section{Check for hyperbolicity}
\label{sec:Check}

A direct check of hyperbolicity of the attractor may be undertaken via the method of calculation of distribution of angles between stable and unstable invariant manifolds of the chaotic saddle trajectories on the attractor \cite{SPKuz1,LGY}. If these angles are strictly non-zero, one can conclude that the attractor is hyperbolic. Otherwise, a non-vanishing probability of the angles in a vicinity of zero value gives an evidence of a non-hyperbolic nature of the attractor (since the probability of tangencies between stable and unstable manifolds remains non-zero). According to the Newhouse lemma, in dissipative systems such tangent situations may lead to appearance of a ``quasi-attractor'' \cite{ABS,Katok}.

In our case, there are 1D unstable manifold and 3D stable manifold associated with any chaotic trajectory in the phase space of the 4D smooth invertible map. The algorithm of calculation of the angles between the manifolds consists in the following. First, we integrate equations~(\ref{eq:e1}) numerically and obtain a sufficiently long trajectory ${\bf Z}(t)=(x(t),y(t),u(t),v(t))$ on the attractor. Second, we integrate the linearized system~(\ref{eq:e6}) forward in time. In the course of integration, normalization of the vector $\tilde{{\bf Z}}^F=(\tilde{x}(t),\tilde{y}(t),\tilde{u}(t),\tilde{v}(t))$ is performed at each step in order to avoid the divergence. Third, we solve a set of 4 systems~(\ref{eq:e6}) with different initial conditions along the same chaotic trajectory but backward in time, so that we obtain a set of vectors $\tilde{{\bf Z}}_k^B$, $k=1 \ldots 4$  of the same kind. In order to avoid dominance of one vector and divergence, we use the Gram-Schmidt procedure of orthogonalization and normalization at each step of integration.

At each point of stroboscopic section of the trajectory at $t_n=t_0+nT$ the vector $\tilde{{\bf Z}}^F(t_n)$ defines the unstable direction which is tangent to the unstable 1D manifold, while the set of first three vectors $\{\tilde{{\bf Z}}_1^B(t_n),\tilde{{\bf Z}}_2^B(t_n),\tilde{{\bf Z}}_3^B(t_n)\}$ defines the basis in the stable 3D subspace, so that any linear combination of these vectors will also belong to the stable subspace. As regards the forth vector $\tilde{{\bf Z}}_4^B$, it is orthogonal to the previous three ones in accordance with the Gram-Schmidt procedure, hence it is orthogonal to the whole stable subspace. In order to define the angle $\alpha$ between stable and unstable subspaces, first let us define the angle $\beta \in [0,\pi/2]$ between the vectors $\tilde{{\bf Z}}^F(t_n)$ and $\tilde{{\bf Z}}_4^B(t_n)$: $\cos{\beta_n}=|\tilde{{\bf Z}}^F(t_n)\tilde{{\bf Z}}_4^B(t_n)|/|\tilde{{\bf Z}}^F(t_n)||\tilde{{\bf Z}}_4^B(t_n)|$. Then we finally get $\alpha_n=\pi/2-\beta_n$.

\begin{figure}[htbp]
\centerline{\includegraphics[width=4.0in]{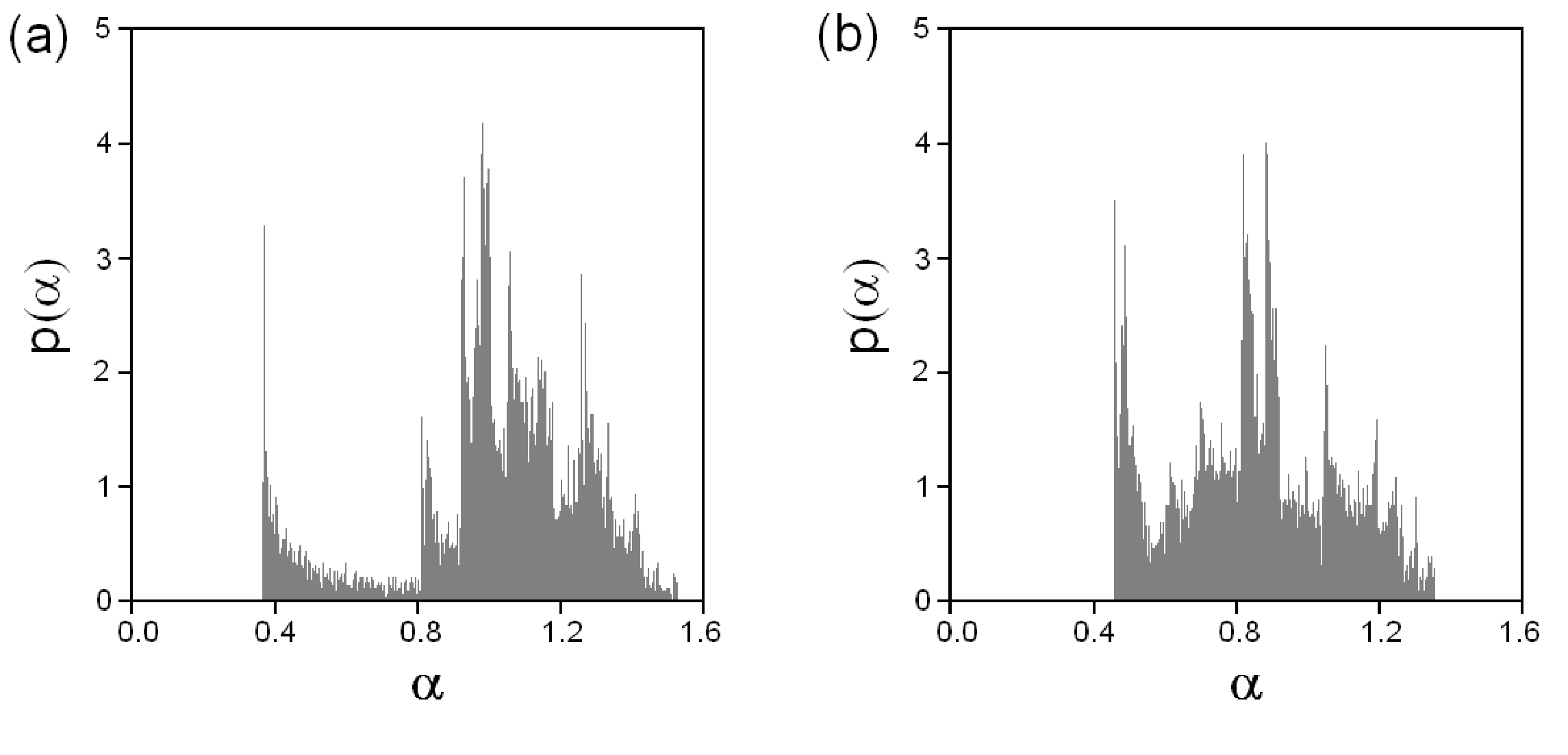}}
\caption{Histograms for distributions of angles $\alpha$ between the stable and unstable subspaces obtained as described in the text at
$A_0=1.5$, $B_0=0.1$, $B_1=0.1$, $c=0.2$, $\Omega=0.05$. Other paraameter values are $A_1=1.7$, $\varepsilon=0.7$ (a) and $A_1=1.5$, $\varepsilon=0.4$ (b).}
\label{fig:f3}
\end{figure}

A histogram of the probability density distribution $p(\alpha)$ for the last parameter values (see comment for Fig.~\ref{fig:f2}) is presented in Fig.~\ref{fig:f3}(a), and Fig.~\ref{fig:f3}(b) shows a histogram for $A_0=1.5$, $A_1=1.5$, $B_0=0.1$, $B_1=0.1$, $c=0.2$, $\Omega=0.05$, $\varepsilon=0.4$. One can see that the distributions are strictly bounded from zero values of angles, so that homoclinic tangencies of manifolds are excluded.

\begin{figure}[htbp]
\centerline{\includegraphics[width=3.5in]{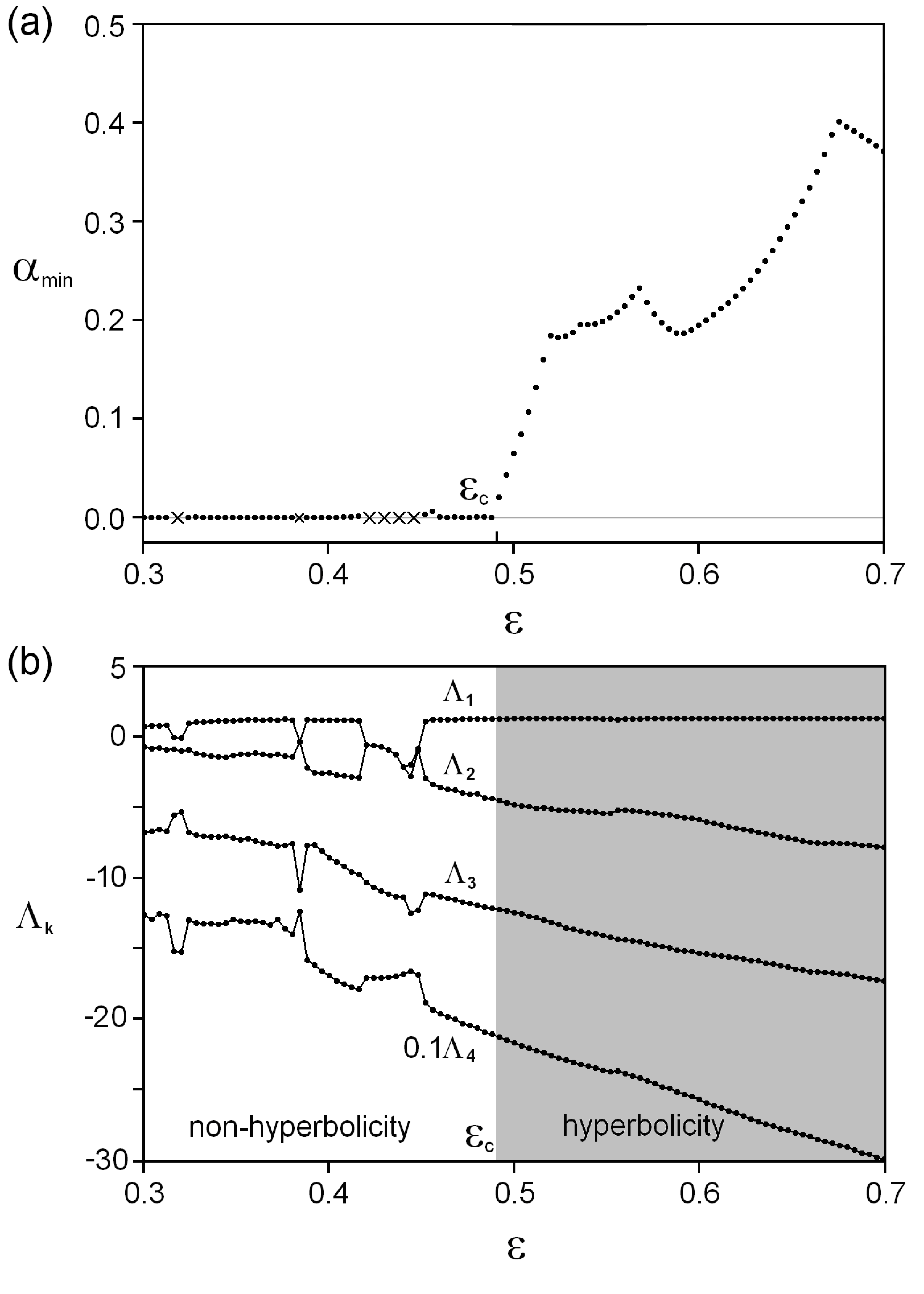}}
\caption{Computed minimum angle $\alpha_{min}$ (a) and Lyapunov exponents (b) of the stroboscopic map~(\ref{eq:e3}) versus parameter $\varepsilon$ at $A_0=1.5$, $A_1=1.7$, $B_0=0.1$, $B_1=0.1$, $c=0.2$, $\Omega=0.05$.}
\label{fig:f4}
\end{figure}

To assess how typically do hyperbolic chaotic regimes occur in intervals of the parameter values of the system~(\ref{eq:e1}), we have obtained a dependence of the minimum angle $\alpha_{min}(N)=\min_{1 \leq n \leq N}\alpha_n$, calculated over a segment of trajectory of length $N=10^4$ points (with random initial conditions), vs. parameter $\varepsilon$ (Fig.~\ref{fig:f4}(a)). In this diagram the values of $\alpha_{min}$ for chaotic regimes are marked by thick dots ($\bullet$), while crosses ($\times$) denote ``gaps'' of regular (periodic) dynamics within the chaotic region, where the angles between manifolds are not defined. One can see, that the diagram can be divided into two intervals. For $\varepsilon<\varepsilon_{c}$, the values of $\alpha_{min}$ are of order $10^{-4}-10^{-3}$, and they tend to zero if the length $N$ is increased. As $\varepsilon$ increases over $\varepsilon_{c}$, the values of $\alpha_{min}$ become sufficiently nonzero, and they do not diminish as the length $N$ rises. Therefore, we can suppose that the chaotic dynamics of the system~(\ref{eq:e1}) for $\varepsilon>\varepsilon_{c}$ becomes hyperbolic. The critical value $\varepsilon_{c}$ was estimated from the analysis of histograms of $\alpha_n$ and values of $\alpha_{min}$ as $\varepsilon_{c}=0.49 \pm 0.01$.

If the attractor is indeed hyperbolic, the chaotic dynamics of the system is robust, i.e. the character of the dynamics should not change under (relatively small) variation of the control  parameters. We have also obtained the dependence of Lyapunov exponents vs. coupling parameter $\varepsilon$, shown in Fig.~\ref{fig:f4}(b). One can see that the parameter range for $\varepsilon$ in this figure may be roughly divided into two intervals. For $\varepsilon<\varepsilon_{c}$ the dependence of Lyapunov exponents vs. $\varepsilon$ is indented, and the value of the exponent varies sufficiently. In this interval the regions of existence of non-hyperbolic chaos alternate with the regions of periodic dynamics. For $\varepsilon>\varepsilon_{c}$ the dynamics of the system changes essentially, and the hyperbolic chaos arises. On can see that within the region of the hyperbolic chaos the largest Lyapunov exponent retains the value $\Lambda_1 \approx \ln{4}$, while the other exponents demonstrate rather smooth (approximately linear) parameter dependencies.

\section{Non-hyperbolic case}
\label{sec:Non-hyp}

\begin{figure}[htbp]
\centerline{\includegraphics[width=6.5in]{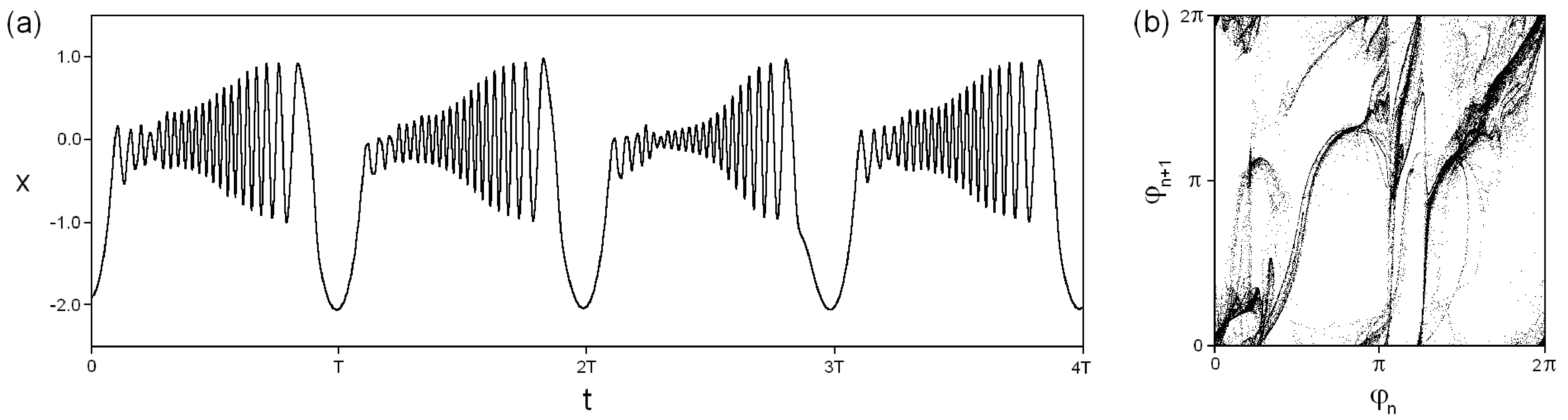}}
\caption{An example of time realization (a) and phase mapping (b) for non-hyperbolic chaotic attractor in the system~(\ref{eq:e1}) at $A_0=1.5$, $A_1=1.7$, $B_0=0.1$, $B_1=0.1$, $c=0.2$, $\Omega=0.05$, $\varepsilon=0.3$.}
\label{fig:f5}
\end{figure}

For comparison, let us consider an example of a non-hyperbolic chaotic attractor in the system~(\ref{eq:e1}) at $A_0=1.5$, $A_1=1.7$, $B_0=0.1$, $B_1=0.1$, $c=0.2$, $\Omega=0.05$, $\varepsilon=0.3$. A fragment of time series for the variable $x$ is shown in the Fig.~\ref{fig:f5}(a), and the diagram for the mapping of the phase $\varphi_{n+1}(\varphi_n)$ is shown in the Fig.~\ref{fig:f5}(b). One can see that the oscillations have well expressed laminar and bursting  phases, as previously.
However, the diagram for the phase variable map does not seem to possess any obvious structure, and it looks typically for quasi-attractors. This conclusion is also confirmed by calculation of Lyapunov exponents of the attractor and by a histogram of the angles between the manifolds. The values of Lyapunov exponents for stroboscopic Poincar\'e map appears to be equal $\Lambda_1 \approx 0.813$, $\Lambda_2 \approx -0.625$, $\Lambda_3 \approx -6.72$, $\Lambda_4 \approx -125.9$.

\begin{figure}[htbp]
\centerline{\includegraphics[width=4.0in]{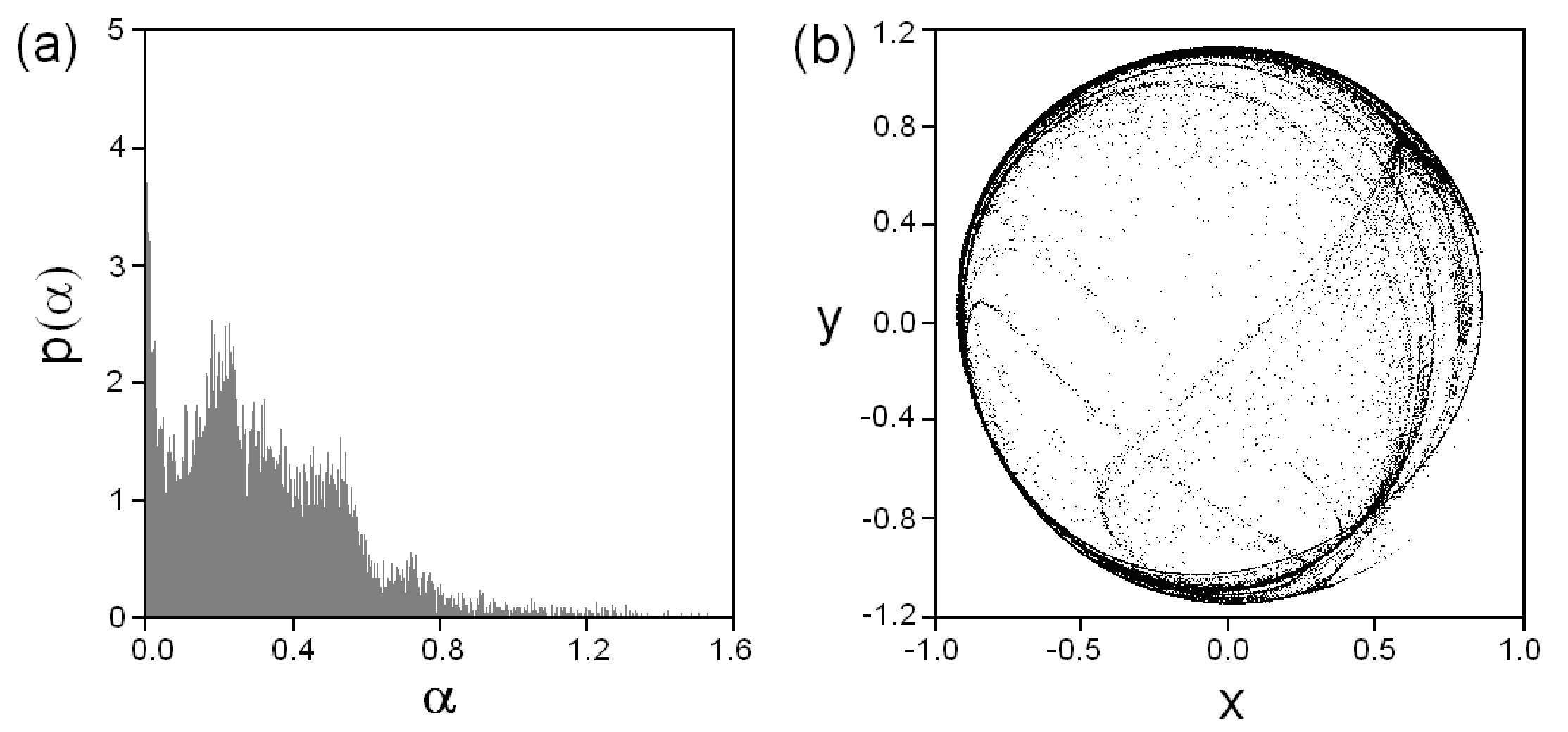}}
\caption{Histogram of distribution of angles between stable and unstable subspases (a) and a portrait of the strange attractor in the stroboscopic Poincar\'e section in projection onto the plane $(x,y)$ (b).}
\label{fig:f6}
\end{figure}

The histogram of angles distribution is shown in Fig.~\ref{fig:f6}(a), which demonstrates that the probability remains non-vanishing nearly the zero value. A projection of a phase portrait in Fig.~\ref{fig:f6}(b) demonstrates a ``diffusive'' set and does not look like a Smale-Williams solenoid, which has a structure of a fractal ``thread'' with infinite number of turns. In the aggregate, all this characteristics testify about the chaotic nature of the dynamics, which is absolutely different from the cases considered previously.

\section{Conclusion}
\label{sec:Conc}

In conclusion note that hyperbolic attractors possessing well-ordered phase space structure and rather simply determined (although chaotic) time dynamics show a very good example of so called ``order in chaos''. It is enough to compare diagrams of mappings for phase dynamics and portraits in the state space for hyperbolic and non-hyperbolic attractors to illustrate this statement. Therefore, a discovery of hyperbolic chaos in the systems of different nature, including neurophysiologic, seems to be important both from the ``world outlook'' and technical viewpoints. Namely, one can expect that complex networks constructed from a large number of neurons may demonstrate special properties of stability with respect to parameters variation and noise, as well as special properties of synchronization and of transitions ``order-to-chaos'', compared with usual chaotic systems.

\section*{Acknowledgment}
The author thanks Prof.~S.~Kuznetsov and Dr.~I.~Sataev for helpful discussion. This work was supported by RFBR grant No 12-02-00342.

\end{document}